\documentclass[11pt,aps,prc,preprint,doublespace,showpacs,
amssymb,byrevtex,superscriptaddress,nofootinbib]{revtex4}  

\usepackage{epsfig,bm,dcolumn}

\begin{document}


\title{Investigating the parity of the exotic $\Theta^+$ baryon from
kaon photoproduction
}


\author{Byung Geel Yu}%

\email{bgyu@mail.hangkong.ac.kr}

\affiliation{Department of General studies,
Hangkong Aviation University, Koyang, 200-1, Korea}
\affiliation{Department of Physics, North Carolina State
University, Raleigh, North Carolina 27695-8202, USA}

\author{Tae Keun Choi}%

\email{tkchoi@dragon.yonsei.ac.kr}
\affiliation{Department of Physics, Yonsei University, Wonju, 220-710, Korea}

\author{Chueng-Ryong Ji}%

\email{crji@unity.ncsu.edu}
\affiliation{Department of Physics, North Carolina State
University, Raleigh, North Carolina 27695-8202, USA}

\date{\today}


\begin{abstract}

Based on the hadronic model with the gauge prescription suggested
by Ohta and Haberzettl, we investigate the possibility of
determining the parity state of the $\Theta^+$ baryon using photon
induced processes, $\gamma n\to K^- \Theta^+$ and $\gamma p\to
\bar{K}^0 \Theta^+$. The total and differential cross sections are
simulated in two versions of pseudovector(PV) and pseudoscalar(PS)
coupling schemes and the results are reported both on the positive
and negative parity states of the $\Theta^+$ baryon. It is found
that in both coupling schemes the total cross sections from the
neutron target are in general larger than those from the proton
target, regardless of the $\Theta^+$ parities. The cross sections
of the $\Theta^+$ production however depend largely on the value
of the $\Theta^+$ decay width which is not yet well established.
Moreover, there is a wide theoretical uncertainty associated with
the different assumption on the gauge prescription in model
calculations. We discuss these points by comparing theoretical
predictions with the existing experimental data. Our analysis
suggests that the observation of the angular distribution rather
than just the total cross section in the photoproduction process
may be a useful tool to distinguish the parity of the $\Theta^+$
baryon.

\end{abstract}


\maketitle

\section{Introduction}

The recent experimental observations of the narrow baryon state
from the invariant mass spectrum of $K^+ n $ or $K^0 p$ in photon
induced nuclear reactions and their interpretation as an exotic
pentaquark state of the $\Theta^+$ baryon with s=+1 attracted a
lot of attention \cite{leps,saph,clasi,clasii,herm}. Such an
experimental evidence for the $\Theta^+$ baryon was also observed
in other reaction channels, $K^+ Xe\to K^0\,p\,Xe^{\prime}$
\cite{dian}, $\nu_{\mu^-}(\bar{\nu}_{\mu^-})$ collisions with
nuclei \cite{asra} and $pA\to pK^0_s X$ \cite{svd}, which tend to
confirm the existence of the $\Theta^+$ baryon. The extracted mass
of the $\Theta^+$ baryon from these experiments is reported to be
1.54 GeV and its decay width less than 25 MeV are consistent with
those of the pentaquark state predicted in the chiral soliton
model \cite{diak,weig,chem,pras}.

These experimental identifications of the $\Theta^+$ have
initiated intensive studies of the new type of hadron structures
that are containing more than two or three quarks
\cite{jaff,jaff01}. However, since quantum numbers  other than its
mass and decay width of the detected $\Theta^+$ baryon are not yet
known from these experiments, much theoretical attention has been
paid to the determination of its further properties
like spin, isospin, parity and its magnetic moment. 
Subsequent theoretical investigations on the structure
of the $\Theta^+$ baryon follow based on the constituent
quark model including diquark-diquark-$\bar{q}$
approach \cite{cheu,stan,carl,golz,will,karl,shur},
Skyrme model \cite{weig,chem,pras,wall,jenn,bori,itzh,wu},
QCD sum rule \cite{zhu,math,sugi},
chiral potential model \cite{hosa}, large $N_c$ QCD \cite{cohe},
lattice QCD \cite{csik,sasa} and Group
theory approach \cite{wybo,bijk}. The dynamical properties of the $\Theta^+$
baryon was also studied through the production of the $\Theta^+$ in the
relativistic nuclear collisions \cite{rand,chen}.

All these theoretical studies address  various aspects of the
$\Theta^+$ baryon properties and in many cases the models assume or predict a
definite parity for the $\Theta^+$
as positive \cite{diak,bori,carl,stan,hosa,golz,karl,math,shur}.
However, recent works from the QCD-sum rules \cite{zhu,sugi} and
the lattice QCD \cite{csik,sasa} favor a negative parity.
Therefore the assumptions on or the model predictions for
the parity of the $\Theta^+$ are still controversial and it is
thus of importance to analyze the processes that may reveal the
true parity state of the $\Theta^+$.

Along this line of thoughts, there are theoretical attempts to
determine the parity of the $\Theta^+$ baryon by the direct
estimation of the cross sections observed in the photon and meson
induced $\Theta^+$ production experiments using hadronic models
\cite{nam,cmko01,ysoh,zhao,naka}. In particular, the cross
sections of $\gamma n\to K^-\Theta^+$ and $\gamma p\to
\bar{K}^0\Theta^+$ have been estimated with the hadronic models
including hadron form factors \cite{cmko01,ysoh} and compared with
the data from the SAPHIR experiment \cite{saph}. The use of hadron
form factors requires, of course, the gauge invariance of the
photoproduction amplitude and this constraint is indeed satisfied
in Refs. \cite{cmko01,ysoh}. Yet, in view of the model development
in the similar processes, $\gamma p\to K^+\Lambda$ and $\gamma
p\to K^+\Sigma^0$ \cite{jans01,jans02,benn,mart}, the gauge
prescription suggested by Ohta \cite{ohta} and Haberzettl
\cite{habe} yields the better $\chi^2$-fit for the analysis of the
$\gamma p\to K^+\Lambda$ process and has a firm field theoretic
foundation. Since this point cannot be overlooked, we apply the
prescription of Refs. \cite{ohta,habe} to the model calculation of
the processes $\gamma n\to K^-\Theta^+$ and $\gamma p\to
\bar{K}^0\Theta^+$.

To focus on the difference in the gauge prescription from the
earlier work, we use the same model parameters of Ref.
\cite{ysoh}. While Ref.\cite{ysoh} concluded that the total cross
section already determined the parity of the $\Theta^+$ as
positive, our results indicate that there is a wide theoretical
uncertainty associated with the different assumptions on the gauge
prescription and the total cross section itself cannot yield a
definite conclusion on the parity of the $\Theta^+$.
We thus stress that further analysis of
the angular distributions is necessary. Especially, the features
of the angular distribution near threshold become less dependent
on the model parameters because they follow the conservation rules
of parity and angular momentum.

This paper is organized as follows. In Sec. II,  the cross section
and the differential cross section for angular distribution are
evaluated for the $\Theta^+$ production from $\gamma N\to
K\Theta^+$ when the $\Theta^+$ has the positive parity. The cross
section and angular distribution of the reaction in the case of
the negative parity $\Theta^+$ production are evaluated in Sec.
III. Summary and discussion follow in Sec. IV.

\section{photoproduction for the positive parity $\Theta^+$}

\begin{figure}[tr]
\centering
\epsfig{file=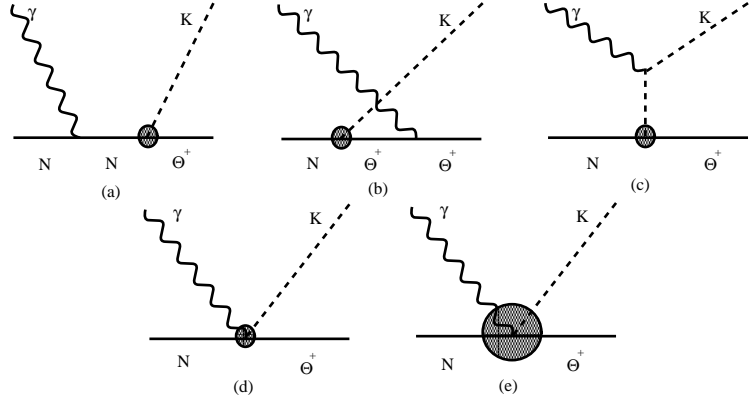, width=0.6\hsize}
\caption{Tree level diagrams for $\gamma N \to K \Theta^+$ reaction. Diagrams (a), (b) and
(c) denote the s-, u- and t-channel pole terms with hadron form factors depicted as
the blob at each vertex. The diagram (d) is
the Kroll-Ruderman(KR) term; it is absent for pseudoscalar couplings with bare vertices.
The last diagram (e) corresponds to the contact interaction term required to restore
gauge invariance of the Born amplitude. It is depicted as a large blob to distinguish from
the KR term.  }
\label{fig:fig01}
\end{figure}

The photoproduction of the $\Theta^+$ baryon from neutron or proton
target is usually calculated in relativistic hadron models because the models
with hadronic degrees of freedom are more relevant than the perturbative QCD
to the energy range of the reactions that we study in this work.
In hadronic models the reaction is
generated from the Feynman diagrams at tree level as shown
in Fig. \ref{fig:fig01}. The momenta of the incident
photon, the nucleon, the outgoing kaon, and the $\Theta^+$ are $k$, $p$,
$q$, and $p'$, respectively in the diagrams of Fig. \ref{fig:fig01}.
The Mandelstam variables are
$s=(p+k)^2$, $t=(k-q)^2$, and $u=(p^{\prime}-k)^2$.
Using effective Lagrangians for vertex couplings pertinent to the
diagrams, the transition amplitude is obtained. Here, as the
interaction Lagrangians relevant to the process are found in many
literatures \cite{cmko01, cmko02,ysoh,naka}  we will not repeat
them. Instead, with the interaction Lagrangians given in Refs.
\cite{cmko01, ysoh}, let us begin with the Born amplitude for the
positive parity $\Theta^+$ photoproduction. The Born amplitude of
the PV coupling $KN\Theta^+$ interaction can be written as
\begin{eqnarray}\label{00}
{\cal M}_{Born}={\cal M}_{PV-pole}+{\cal M}_{KR}+{\cal M}_{c}\,.
\end{eqnarray}
The PV-pole terms are composed of the first three pole diagrams of (a), (b) and
(c) in Fig. \ref{fig:fig01} which correspond to the nucleon, $\Theta^+$ and
kaon exchanges in the s-, u- and t-channel respectively, i.e.,
\begin{eqnarray}\label{posi-pv-pole-01}
&&{\cal M}_{PV-pole}\nonumber\\&&
=\frac{e\, g_{KN\Theta}}{M+M_{\Theta}}\bar{u}_{\Theta}(p^{\prime})\biggl\{
F_{1}(s)\gamma_{5}/\kern-5pt{q}\frac{(/\kern-6pt{p}+/\kern-5pt{k}+M)}{s-M^{2}}\,
[Q_{N}\,/\kern-5pt{\epsilon} +i\frac{\kappa_{N}}{2M}
\sigma^{\mu\nu}\epsilon_{\mu}k_{\nu}]
+[Q_{\Theta}\,/\kern-5pt{\epsilon}
+i\frac{\kappa_{\Theta}}{2M_{\Theta}}\sigma^{\mu\nu}\epsilon_{\mu}k_{\nu}]\nonumber\\&&\times
 \frac{(/\kern-6pt{p^{\prime}}-/\kern-5pt{k}
+M_{\Theta})}{u-{{M_{\Theta}}^{2}}}\gamma_{5}/\kern-5pt{q}F_{2}(u)
+Q_K
F_{3}(t)\gamma_{5}(M+M_{\Theta})\frac{(2q-k)\cdot\epsilon}{t-m_{K}^{2}}\,
\biggr\}\,u_{N}(p).
\end{eqnarray}
For brevity,  we write the amplitude collectively for $\gamma n\to
K^-\Theta^+$ and $\gamma p\to \bar{K}^0\Theta^+$ with notations
$Q_N$, $Q_\Theta$ and $Q_K$ by assigning $Q_N=0$, $Q_{\Theta}=1$
and $Q_K=-1$ to $\gamma n\to K^-\Theta^+ $, and $Q_N=1$,
$Q_{\Theta}=1$ and $Q_K=0$ to $\gamma p\to \bar{K}^0\Theta^+ $,
respectively. The anomalous magnetic moments of the proton and
neutron are $\kappa_p=1.79$ and $\kappa_n= -1.91$. In
Eq.(\ref{posi-pv-pole-01}), $g_{KN\Theta}$ is the $\Theta^+$
coupling constant, $\kappa_\Theta$ is the anomalous magnetic
moment of $\Theta^+$, and $\epsilon$ is the photon polarization
vector. Also, $F_{1}(s)$=$F_{1}(s, {M^{\prime}}^2, m_{\pi}^2)$,
$F_{2}(u)$=$F_{2}(M^2, u, m_{\pi}^2)$ and $F_{3}(t)$=$F_{3}(M^2,
{M^{\prime}}^2, t)$ are the hadron form factors introduced to
the $KN\Theta$ vertices in the s-,u- and t-channel with the
normalizations $F_{1}(s=M^2)=1$, $F_{2}(u={M^{\prime}}^2)=1$ and
$F_{3}(t=m_{\pi}^2)=1$, respectively.

In the PV coupling, the
Kroll-Ruderman term of Fig. \ref{fig:fig01} (d) is required to
restore gauge invariance
of PV pole terms 
due to the
$\gamma_5/\kern-5pt{q}$ coupling;
\begin{eqnarray}\label{KR}
{\cal M}_{KR}=-\frac{e\,g_{KN\Theta}}{M+M_{\Theta}}
\bar{u}_{\Theta}(p^{\prime})\,\gamma_5\,/\kern-5pt{\epsilon}
\biggl\{\,F_{1}(s)\,Q_N- Q_{\Theta}\,F_{2}(u)  \biggr\}\,u_{N}(p).
\end{eqnarray}
In the gauge transformation of the PV coupling pole terms together
with the Kroll-Ruderman term, however, these amplitudes are not
gauge invariant, but yield the following relation,
\begin{eqnarray}\label{ward}
({\cal M}_{PV-pole}+{\cal
M}_{KR})_{\epsilon\to
k}=e\,g_{KN\Theta}\,\bar{u}_{\Theta}(p^{\prime})\gamma_{5}\biggl\{
F_{1}(s)\,Q_{N}\,-Q_{\Theta}\,F_{2}(u)\,-Q_K\,F_{3}(t)\biggr\}
u_N(p).
\end{eqnarray}
The nonvanishing divergence of Eq.(\ref{ward}) is due to the
use of different form factor for each hadron vertex 
and this sort of
divergence equally holds for the gauge transformation of PS
coupling photoproduction amplitude. In fact it vanishes when
$F_i=1$ for $i=1,2,3$, i.e., in the case of point interaction of
$KN\Theta$, or when $F_i=F$ for all $i$, i.e., in the case of
using an overall form factor, $F$. In Refs. \cite{cmko01,ysoh}, the
recipe they used in order to restore gauge invariance of the Born
amplitude as given in Eq.(\ref{ward}) corresponds to the case of
using an overall form factor. In this work, we follow the
gauge prescription suggested by Ohta \cite{ohta} and later improved further
by Haberzettl \cite{habe}. According to these field theoretic
analyses \cite{ohta,habe}, the divergence of the hadronic current due
to the different form factors can be removed by introducing the
diagram (e) of Fig. \ref{fig:fig01}, so called contact interaction
term. It is of the form;
\begin{eqnarray}\label{oh}
{\cal
M}_c&=&-e\,g_{KN\Theta}\,\bar{u}_{\Theta}(p^{\prime})\gamma_{5}\biggl\{
(F_{1}(s)-\widehat{F})\,Q_{N}\frac{(2p+k)\cdot \epsilon}{s-M^{2}}\,
+Q_{\Theta}\,(F_{2}(u)-\widehat{F})\frac{(2p^{\prime}-k)\cdot
\epsilon}{u-{{M_{\Theta}}^{2}}}\nonumber\\&&
+Q_K(F_{3}(t)-\widehat{F})\frac{(2q-k)\cdot \epsilon}{t-m_{K}^{2}}
\biggr\} u_N(p).
\end{eqnarray}
Here, $\widehat{F}$ is a subtraction function which depends on the
Mandelstam variables $(s,u,t)$. Note that in order to maintain the
original singularity structure of the Born amplitude, each of the
three pole terms in Eq.(\ref{oh}) should be nonsingular, i.e.,
$\widehat{F}=1$ for on-mass shell and this can be a constraint on the
arbitrary choice of the function $\widehat{F}$ \cite{habe01, davi}. In this
work, to preserve the crossing symmetry of the amplitude, we
choose the subtraction function, in specific,
\begin{eqnarray}\label{dw}
\widehat{F}=F_{2}(u)+F_{3}(t)-F_{2}(u)F_{3}(t), ~~~~~
\widehat{F}=F_{1}(s)+F_{2}(u)-F_{1}(s)F_{2}(u)
\end{eqnarray}
for $\gamma n \to K^-\Theta^+$ and $\gamma p \to \bar{K}^0\Theta^+$, respectively.
For each hadron form factor in the channels $x=s, u, t$,(or $i$=1, 2, 3), we use
\begin{equation}
F_i(x,M_i) = \frac{\Lambda^4}{\Lambda^4 + (x -
M_i^2)^2}\,,
\label{ff}
\end{equation}
which are normalized to unity at $x= M_i^2$.
Here $M_i$ is the mass of the exchanged particle and $x$ is the
square of the transferred momentum. This function has the
correct on-shell condition, i.e., $F_i(x=M_i^2)=1$ for $i$=1, 2, 3
and, thus, $\widehat{F}=1$ by Eq.(\ref{dw}).

In the PS coupling, the Born amplitude is composed of those terms
depicted by Figs. \ref{fig:fig01}(a) - (c). By the procedure
similar to that of PV coupling, the Born amplitude which preserves
gauge invariance can be obtained by
\begin{eqnarray}\label{ps00}
{\cal M}_{Born}={\cal M}_{PS-pole}+{\cal M}_{c}\,,
\end{eqnarray}
where
\begin{eqnarray}
&&{\cal M}_{PS-pole}\nonumber\\&&=e\,g_{KN\Theta}\,\bar{u}_{\Theta}(p^{\prime})\biggl\{
F_{1}(s)\,\gamma_5\,\frac{(/\kern-6pt{p}+/\kern-5pt{k}+M)}{s-M^{2}}\,
[\,Q_{N}\,/\kern-5pt{\epsilon} +i\frac{\kappa_{N}}{2M}
\sigma^{\mu\nu}\epsilon_{\mu}k_{\nu}\,]
+[\,Q_{\Theta}\,/\kern-5pt{\epsilon}
+i\frac{\kappa_{\Theta}}{2M_{\Theta}}\sigma^{\mu\nu}\epsilon_{\mu}k_{\nu}\,]\nonumber\\&&\times
 \frac{(/\kern-6pt{p^{\prime}}-/\kern-5pt{k}
+M_{\Theta})}{u-{{M_{\Theta}}^{2}}}\,\gamma_{5}\,F_{2}(u)
+Q_K\,F_{3}(t)\gamma_{5}\frac{(2q-k)\cdot\epsilon}{t-m_{K}^{2}}\,
\biggr\}\,u_{N}(p),
\end{eqnarray}
and the contact interaction term of Fig. \ref{fig:fig01} (e) is
given by Eq.(\ref{oh}).

In the calculation of cross sections based on this framework, the
coupling constant $g_{KN\Theta}$ and  the anomalous magnetic
moment $\kappa_{\Theta}$  are to be determined. Unfortunately,
there are no detailed informations available on these quantities
at present. Instead, we have only  few experimental observations;
the decay width $\Gamma_\Theta$ and the cross section. It has been
reported that the decay width $\Gamma_{\Theta}$ is measured in the
range $9\sim 25$ MeV \cite{leps, saph,clasi,clasii,dian,herm,svd}
and the mean cross section for $\gamma p\to \bar{K}^0\Theta^+$ in
the SAPHIR experiment is in the order of 200 nb up to $E_\gamma=$
2.6 GeV \cite{saph}. More recently HERMES experiment estimated the
cross section of the $\Theta^+$ production  to be 100 $\sim$ 220
nb from the quasi-real photoproduction on deuteron, $eD\to pK^0_s
X$ \cite{herm}. However, the precise measurements of the width and
cross sections are still lacking and there are on-going
discussions about possible reanalyses of these observables
\cite{nuss,arnd,clos}. In particular, using the $K^+ d$ scattering
data, Nussinov reanalyzed the decay width of $\Theta^+$
and came up with $\Gamma_{\Theta}< 6$ MeV \cite{nuss}. Moreover, Arndt {\it
et. al.} suggested $\Gamma_{\Theta}\leq 1$ MeV based on the $K^+ p$
and $K^+ d$ database \cite{arnd}. In this work, we adopt
$g_{KN\Theta}$=2.2 assuming $\Gamma_\Theta \simeq 5$ MeV for the
positive parity $\Theta^+$ and compare our results to the present SAPHIR
data \cite{saph}. The value of $\kappa_\Theta$ is still
elusive, although there are some theoretical suggestions on this
quantity \cite{diak,jaff}. We consider it as a parameter and vary
its value between $-0.7 \leq \kappa_\Theta \leq 0.7$.
\begin{figure}[tr]
\centering \epsfig{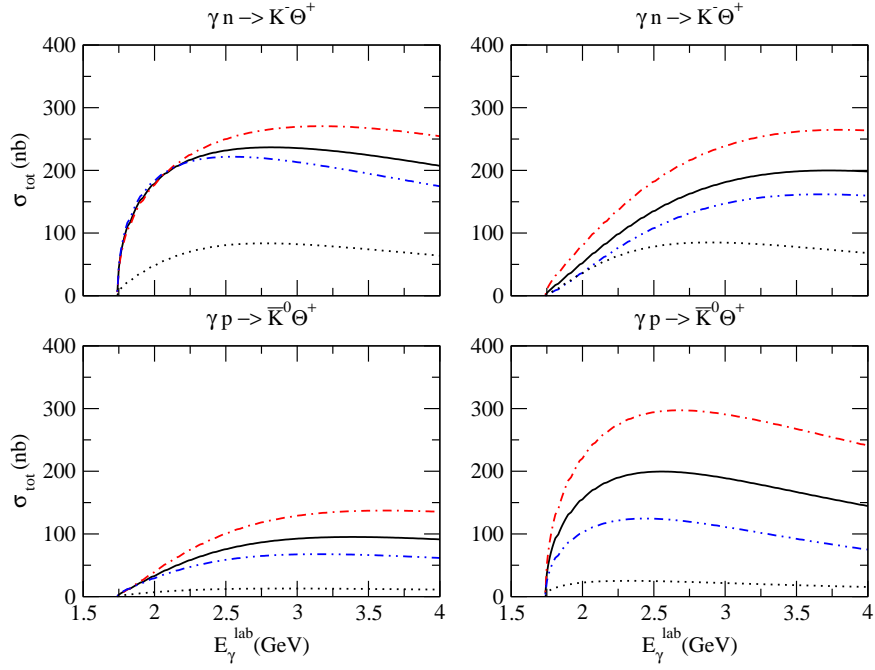} \caption{Cross
sections for $\gamma n \to K^- \Theta^+$ and for $\gamma p \to
\bar{K}^0 \Theta^+$ when the $\Theta^+$ has positive parity. The
PV coupling scheme is displayed in the left column and the PS
scheme is in the right column.
Given the coupling constant $g_{KN\Theta}=2.2$, dependence of the
cross sections with $\Lambda=1.8$ GeV and 1.2 GeV are shown. The
dotted lines are the results of the Born amplitude with
$\kappa_{\Theta}=0$ and cutoff $\Lambda=1.2$ GeV. The solid lines
are the results of the Born amplitude with $\kappa_{\Theta}=0$ and
cutoff $\Lambda=1.8$ GeV. The dot-dashed lines are the Born
contributions with $\kappa_{\Theta}=0.7$ and $\Lambda=1.8$ GeV.
The dot-dot-dashed lines with $\kappa_{\Theta}=-0.7$ and
$\Lambda=1.8$ GeV. } \label{fig:fig02}
\end{figure}
\begin{figure}[tr]
\centering \epsfig{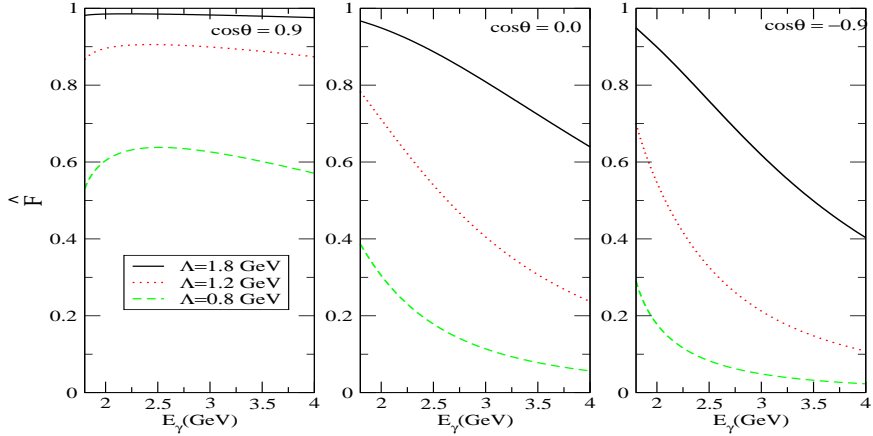} \caption{Energy and angle dependence of the
subtraction function $\hat F$ for $\gamma p\to \bar{K}^0\Theta^+$.
The functional form of $\hat F$ is given by Eq.(\ref{dw}). The
solid line is for $\Lambda=1.8$ GeV. The dotted line and the
dashed line are for $\Lambda=1.2$ GeV and 0.8 GeV respectively.
Note that $\hat{F}=1$ at $E_{\gamma}=0$ below the threshold of the reaction
regardless of $\Lambda$ values.
} \label{fig:fig021}
\end{figure}

The results are given in Fig. \ref{fig:fig02}, where the cross
sections are obtained by using the subtraction function and form
factors of Eqs.(\ref{dw}) and (\ref{ff}) to reduce the strength of
the Born terms. In relation with these functions we present the
sensitivity of the cross sections to the cutoff parameter
$\Lambda$ by taking both $\Lambda=1.8$ GeV \cite{jans01,ysoh} and
a somewhat lower value $\Lambda=1.2$ GeV for comparison. Given the
coupling constant $g_{KN\Theta}=2.2$ with $\kappa_\Theta=0$, the
dotted lines with $\Lambda=1.2$ GeV lower the cross section down
by more than one third of its magnitude as compared to the solid
lines with $\Lambda=1.8$ GeV in Fig. \ref{fig:fig02}. As indicated
in Refs. \cite{jans01,jans02}, it probably makes more sense to
consider the product $g_{KN\Theta}\,F_i(x)$ as the effective
coupling strength but not the bare coupling constant
$g_{KN\Theta}$ alone when form factors are incorporated. For
instance, the effective coupling strength becomes
$g_{KN\Theta}\,F_1(s)\simeq 0.36$ at threshold if $\Lambda=1.2$
GeV. The smaller the cutoff $\Lambda$ is, the more
significantly $\hat{F}$ is attenuated by the reductions in each
form factor $F_i$ as shown in Fig. \ref{fig:fig021}. However, such
a significant fall-off in $\hat F$ may not be so desirable in
order to minimize the ambiguity from the form factors. In Fig.
\ref{fig:fig021}, $\hat F$ is very close to $1$ near threshold
almost independent of the scattering angle $\theta$ if
$\Lambda=1.8$ GeV.
For this reason, we take the $\Lambda=1.8$
GeV in what follows.

In Fig. \ref{fig:fig02}, the cross section for $\gamma n\to
K^-\Theta^+$ is about 150 nb in the PS and 250 nb in the PV scheme
near $E_{\gamma}=2.5$ GeV. The cross section for $\gamma p\to
\bar{K}^0\Theta^+$ is about 200 nb in PS and 80 nb in the PV
scheme. These results are from the Born contributions only and
this point is in sharp contrast to the results of previous
calculations. In Ref. \cite{cmko01}, with more contributions of
the two and three body final state interactions considered, the
authors used the t-channel and u-channel Born terms to obtain
cross sections with the magnitude of 38 nb for $\gamma p\to
\bar{K}^0\Theta^+$ and of 280 nb for $\gamma n\to K^-\Theta^+$ in
the PS scheme. Also in Ref. \cite{ysoh}, the authors included
$K^*$ exchange in their PS coupling Born amplitude to obtain the
cross sections about 320 $\sim$ 400 nb for $\gamma p\to
\bar{K}^0\Theta^+$ and about 200 $\sim$ 230 nb for $\gamma n\to
K^-\Theta^+$, depending on the sign of
$g_{K^*N\Theta}$. But the Born contributions to these total cross
sections are found to be about 40 nb and 80 nb for each process
and almost the rest of the cross sections are from $K^*$
contributions. In fact their Born contributions up to
$E_{\gamma}=4$ GeV are smaller by a factor of $\frac{1}{3}$ or
$\frac{1}{4}$ than our result in PS scheme of Fig.
\ref{fig:fig02},
despite the same cutoff $\Lambda$ with ours. As a consequence, the
$K^*$ contributions relative to the Born terms in the cross
sections are much larger than ours. Furthermore, in contrast to
their findings in the $\kappa_\Theta$ contributions, our cross
sections are significantly dependent on the variation of
$\kappa_{\Theta}$ in case of PS coupling scheme, albeit
parameterized as the same value with Ref. \cite{ysoh}. Besides the
different type of form factor used in Ref. \cite{cmko01} from ours
and Ref. \cite{ysoh}, the apparent distinctions between these
previous results and ours are mainly due to the different gauge
prescriptions adopted in each model calculation. Although neither
of the procedures adopted in Refs. \cite{cmko01,ysoh} violates the
gauge invariance, we emphasize that they certainly need further
improvement in going beyond just taking a single overall form
factor from a field theoretic point of view. In Fig.
\ref{fig:fig02}, it is instructive to note that the cross section
of $\gamma n\to K^-\Theta^+$ near threshold is similar to that of
$\gamma n\to \pi^-p$ and also that of $\gamma p\to
\bar{K}^0\Theta^+$ to $\gamma p\to \pi^0p$, since the two channels
of $\Theta^+$ production have the same charge exchange structure
of the Born terms with the corresponding two processes in the pion
photoproduction. According to the results of Refs. \cite{ysoh01,
diak01} where the couplings of the baryon octet with the
anti-decuplet are analyzed, the $\Theta^+$ is difficult to couple
to any would-be nucleon resonances. We, thus, refer a qualitative
analysis of our cross sections to those of the pion
photoproduction near threshold where no significant contributions
are attributed to the resonances \cite{tiat}. With these in mind,
the "nose" structure of PV coupling scheme of $\gamma n\to
K^-\Theta^+$ near threshold is understood by the Kroll-Ruderman
term and possibly the kaon pole term. The Kroll-Ruderman term
dictates the threshold amplitude, giving large s-wave contribution
to yield a rapid increase of cross section together with the kaon
pole term. For the process $\gamma p\to \bar{K}^0\Theta^+$, there
is neither Kroll-Ruderman term nor kaon pole term due to the
charge conservation. Therefore, the cross section of the latter
process  is suppressed near threshold similar to the case of
$\gamma p\to \pi^0 p$ \cite{hans,fuch}. These qualitative features
are apparent in the PV coupling scheme and consistent with the
remark in Ref. \cite{poly03} that the photoexcitation of the
baryon anti-decuplet is strongly suppressed in the proton target
and the process occurs mostly in the neutron target.

\begin{figure}[tr]
\centering
\epsfig{file=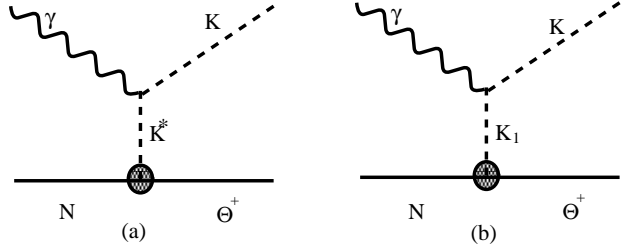, width=0.5\hsize}
\caption{Diagrams for t-channel $K^*$ and $K_1$ exchanges in the $\gamma N \to K \Theta^+$
process.}
\label{fig:fig03}
\end{figure}

We now consider the contributions of t-channel vector meson $K^*$
and $K_1$ axial vector meson exchanges. Fig. \ref{fig:fig03}
depicts the Feynman diagrams for the $K^*$ and $K_1$ exchanges in
the t-channel.
For the $K^*(890)(J^P=1^-)$ exchange, we use the Lagrangians
\begin{eqnarray}
{\cal L}_{K^* N\Theta} &=& g_{K^*N\Theta}\,\bar{\Theta}\biggl(\gamma^\mu  +\frac{\kappa^{*}}{M+M_\Theta}
\sigma^{\nu\mu}\partial_\nu \biggr)K_{\mu}^{*\dagger}\, N + \mbox{h.c.}, \nonumber \\
{\cal L}_{K^* K\gamma} &=& \frac{g_{K^* K\gamma}}{m}\epsilon_{\alpha\beta\mu\nu}
\partial^\alpha A^\beta \partial^\mu K^\dagger K^{*\nu} +\mbox{h.c.},
\label{posi-ks}
\end{eqnarray}
where $g_{K^*N\Theta}$ and $\kappa^*$ are respectively the vector coupling constant and
the tensor coupling ratio of $K^* N\Theta$ vertex. Here $m$ is a parameter of mass dimension
for the anomalous coupling of $g_{K^* K\gamma }$.
The transition amplitude for
$K^*$ exchange in the t-channel can be written as
\begin{eqnarray}
\label{posi-vec}
{\cal{M}}_{K^*}=-G_{K^* N\Theta}\,\bar{u}_{\Theta}\biggl\{ \epsilon_{\alpha\beta\tau\sigma}
k^\alpha\epsilon^\beta q^\tau  \frac{(-g^{\sigma\mu}+
{q^{\prime}}^{\sigma}{q^{\prime}}^{\mu}/M_{K^*}^2)}{t-M_{K^*}^2+i \Gamma M_{K^*}}
\biggl( \gamma_\mu +   i\frac{\kappa^*}{M+M_\Theta} \sigma_{\nu\mu} {q^{\prime}}^{\nu}
   \biggr)\biggr\}u_N\,,
\end{eqnarray}
with $G_{K^* N\Theta}=g_{K^* N\Theta} g_{K^* K\gamma } \,F_3(t)\,
m^{-1}$, and $q_{\mu}^{\prime}=(q-k)_{\mu}$. Including these
contributions, we use $g_{K^*K^{\pm}\gamma}=0.254$ for the charged
kaon anomalous decay and $g_{K^*K^{0}\gamma}=0.388$ for the
neutral kaon decay that are cited in PDG \cite{pdg}. Following
Ref. \cite{cmko02}, the unknown coupling $g_{K^* N\Theta}$ was
deduced to be 1.32 from the assumption, $g_{K^*
N\Theta}/g_{KN\Theta}=0.6$. We adopt this value of $g_{K^*
N\Theta}$ and do not consider the tensor coupling contributions
of both $K^*$ and $K_1$ to avoid any further parameters.
\begin{figure}[tr]
\centering
\epsfig{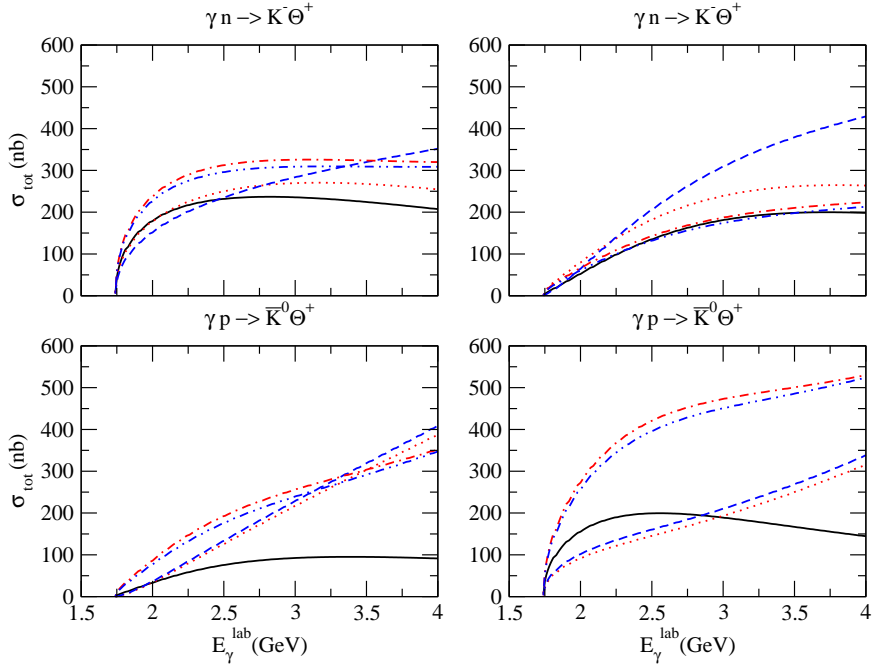}
\caption{Cross sections for $\gamma n \to K^- \Theta^+$ of PV(upper left)
and PS scheme(upper right). Cross sections for $\gamma p \to \bar{K}^0 \Theta^+$ of PV(lower left)
and PS scheme(lower right) when the $\Theta^+$ has positive-parity.
$\kappa_\Theta=0$ in any cases for all panels.
The solid lines are the contributions of the Born amplitude with $g_{K^*N\Theta}=0$.
The dotted lines are the sum of the Born terms and $K^*$  with $g_{K^*N\Theta}= 1.32$.
The dot-dashed lines the sum of the Born terms and $K^*$ with $g_{K^*N\Theta} = -1.32$.
The dashed lines are the sum in total of the Born terms, $K^*$ and $K_1$ with $g_{K^*N\Theta}= 1.32$,
$g_{K_1 N\Theta}= -0.07$ for $\gamma n\to K^-\Theta^+$ and $g_{K_1 N\Theta}= -0.1$ for
$\gamma p\to \bar{K}^{0}\Theta^+$ respectively.
The dot-dot-dashed lines are the sum in total of the Born terms, $K^*$ and $K_1$ with $g_{K^*N\Theta}= -1.32$,
$g_{K_1 N\Theta}= +0.07$ for $\gamma n\to K^-\Theta^+$ and $g_{K_1 N\Theta}= +0.1$ for
$\gamma p\to \bar{K}^{0}\Theta^+$ respectively.
}
\label{fig:fig04}
\end{figure}
The interaction Lagrangians for the axial vector meson $K_1(1270)(J^P=1^+)$ coupling to
$K_1 N\Theta^+$ and $K_1 K\gamma$ are given by
\begin{eqnarray}
{\cal L}_{K_1 N\Theta} &=&  g_{K_1 N\Theta}\,\bar{\Theta}\biggl( \gamma_{\mu}  +
     \frac{\kappa_{1}}{M+M_\Theta}
\sigma_{\nu\mu}\partial^\nu \biggr)K_1^{\mu\dagger}\,  \gamma_5\, N + \mbox{h.c.}, \nonumber \\
{\cal L}_{K_1 K\gamma} &=& -i\frac{g_{ K_1 K\gamma}}{m} K^\dagger (\partial_\mu A_\nu \partial^\mu K_1^\nu -
 \partial_\mu A_\nu \partial^\nu K_1^\mu ) +  \mbox{h.c.},
\label{ax00}
\end{eqnarray}
where $g_{K_1 N\Theta}$ and $\kappa_{1}$ are the axial vector coupling constant and the
tensor coupling ratio of $K_1 N \Theta $ vertex, respectively.
Then, the transition amplitude for the t-channel $K_1$ exchange is given by,
\begin{eqnarray}
\label{posi-ax}
{\cal M}_{K_1}=G_{K_1 N\Theta}\,\bar{u}_{\Theta}
(k\cdot q^{\prime} \epsilon_\mu-\epsilon\cdot q^{\prime} k_\mu)
\frac{(-g^{\mu\nu}+{q^{\prime}}^{\mu}{q^{\prime}}^\nu/M_{K_1}^2)}{t-{M_{K_1}^2}+i \Gamma M_{K_1}}
\biggl(\gamma_\nu +i \frac{\kappa_{1}}{M+M_\Theta}
\sigma_{\alpha\nu} {q^{\prime}}^\alpha\biggr)\gamma_5 u_N,
\end{eqnarray}
with $G_{K_1 N\Theta}=g_{K_1 N\Theta}g_{K_1 K\gamma }\,F_3(t)\,
m^{-1}$. For the axial vector meson coupling constant $g_{K_1
K\gamma}$, there are no empirical data available for the decay
$K_1 \to K \gamma$ except for its decay channel to $\rho$ meson
via the process $K_1\to \rho K$ \cite{pdg}.
\begin{figure}[tr]
\centering
\epsfig{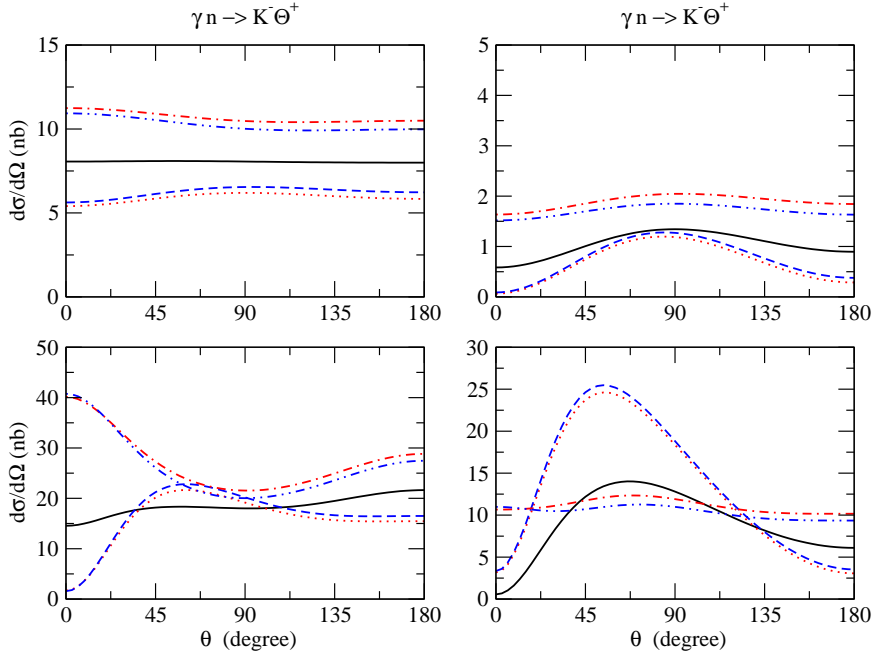}
\caption{Angular distributions for $\gamma n \to K^- \Theta^+$
at $E_\gamma=1.8$ GeV(upper left), $E_\gamma=2.5$ GeV(lower left) of PV
and $E_\gamma=1.8$ GeV(upper right), $E_\gamma=2.5$ GeV(lower right) of PS scheme
when the parity of $\Theta^+$ is positive.
The notations are the same as in Fig. \ref{fig:fig04}}
\label{fig:fig05}
\end{figure}
\begin{figure}[tr]
\centering
\epsfig{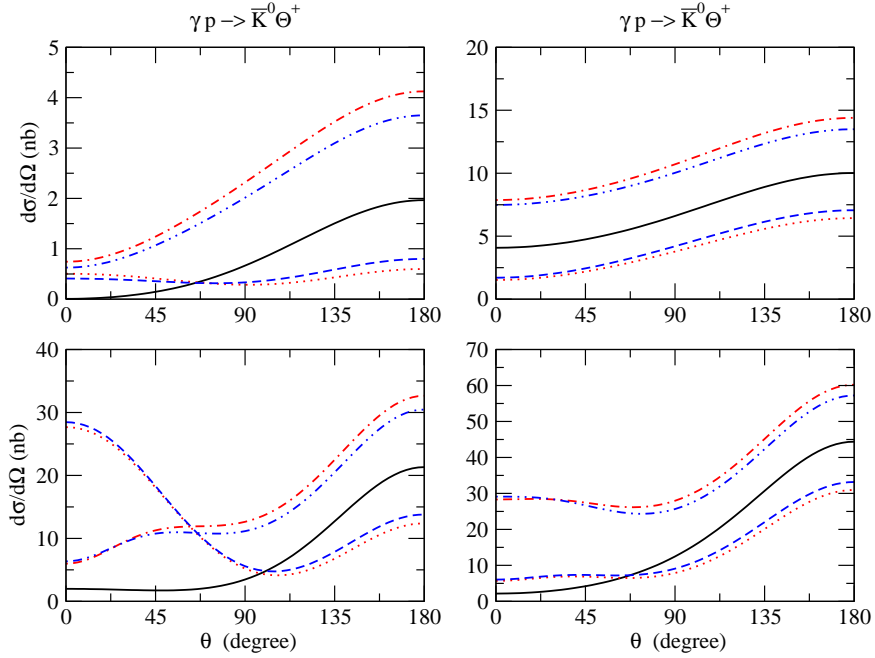}
\caption{Angular distributions for $\gamma p \to \bar{K}^0 \Theta^+$
at $E_\gamma=1.8$ GeV(upper left), $E_\gamma=2.5$ GeV(lower left) of PV
and $E_\gamma=1.8$ GeV(upper right), $E_\gamma=2.5$ GeV(lower right) of PS scheme
when the parity of $\Theta^+$ is positive.
The notations are the same as in Fig. \ref{fig:fig04}}
\label{fig:fig06}
\end{figure}
In Ref. \cite{hagl},
by using the effective Lagrangian given by Eq.(\ref{ax00}) for the
interaction vertex $K_1 K\rho$,\, the decay width $\Gamma_{K_1\to
K\rho }$ was estimated to be 37.8 MeV and the coupling constant
$g_{K_1 K\rho }$ was determined to be 12.0.
Using this value for $g_{K_1 K\rho}$, we deduce the coupling
constant $g_{K_1 K\gamma}=0.6$ by applying the
vector dominance relation for $g_{K_1 K\gamma }$=
$\frac{e}{f_{\rho}}g_{K_1 K\rho}$, where $f_{\rho}^2/4\pi=2.9$. In order
to determine the axial vector coupling constant $g_{K_1 N\Theta}$,
we make use of the ratio $g_{K^*K\gamma}\,g_{K^*p\Lambda}/g_{K_1
K\gamma}\,g_{K_1 p\Lambda}\simeq -8.6$, which is extracted from
WJC model for $K^+\Lambda$ electromagnetic production \cite{crji}
and assume that this ratio is valid also for
$g_{K^*K\gamma}g_{K^*N\Theta}/g_{K_1 K\gamma}g_{K_1 N\Theta}$.
Then, we obtain $g_{K_1 N\Theta}=-0.07$ for $\gamma n\to
K^-\Theta^+$ and $g_{K_1 N\Theta}=-0.1$ for $\gamma p\to
\bar{K}^0\Theta^+$, respectively.
%
%

In Fig. \ref{fig:fig04}, we present the results of $K^*$ and $K_1$
contributions to the cross sections of $\gamma n\to K^-\Theta^+$
and $\gamma p\to \bar{K}^0\Theta^+$ in both coupling schemes. In
most cases, $K^*$ gives significant contributions, whereas the
role of $K_1$ is minor and these vector mesons give contributions
to $\gamma p\to\bar{K}^0\Theta^+$ larger than $\gamma n\to K^-
\Theta^+$. In this figure, the results of these contributions are
more favorable when $g_{K^*N\Theta}=1.32$, $g_{K_1
N\Theta}=-0.07(-0.1)$ for $\gamma n\to K^-\Theta^+$($\gamma p\to \bar{K}^{0}\Theta^+$)
as depicted by the dashed lines. They give the same
order of magnitude to the cross sections for each process at
$E_{\gamma}=2.5$ GeV, regardless of the coupling scheme. Around
this energy, the dashed lines of the cross section for $\gamma
n\to K^-\Theta^+$ are about 230 nb, and for $\gamma p \to
\bar{K}^0\Theta^+$ about 150 nb, respectively.

In Fig. \ref{fig:fig05}, the differential cross sections for
$\gamma n\to K^-\Theta^+$ are displayed near threshold
$E_\gamma=1.8$ GeV and at $E_\gamma=2.5$ GeV. It is interesting to
see that the angular distributions of the kaon produced near
threshold, $E_{\gamma}=1.8$ GeV, are isotropic both in the PV and
PS schemes. These are due to the s-wave dominance from the
Kroll-Ruderman term in the case of PV, and from the s-channel
nucleon pole term in the case of PS, respectively. Notice that the
scales of the cross sections in the two schemes are different.
This feature of the s-wave production near threshold can be
anticipated from the parity and angular momentum conservation
which states that the angular momentum of the produced kaon is in
the s-wave state near threshold, if the parity of the $\Theta^+$
is positive.
In the energy bin $E_\gamma=2.5$ GeV, the interference of $K^*$
and $K_1$ exchanges develops a peak around $\theta=$ 45$^\circ$ in
both coupling schemes when $g_{K^*N\Theta}=1.32$, $g_{K_1
N\Theta}=-0.07$  as depicted by the dashed lines.

In Fig. \ref{fig:fig06}, the angular distributions of $\gamma p\to
\bar{K}^0\Theta^+$ are presented. There appear small deviations
from the isotropic angular distribution in the Born contribution
and the backward asymmetries are observed near threshold. These
backward enhancements hold even at $E_\gamma=2.5$ GeV both in the
PV and PS schemes. They are resulting from the u-channel
contribution of the Born terms, since there is no kaon pole term
in this process. At $E_\gamma=2.5$ GeV, the
change in the sign of $K^*$ and $K_1$ coupling constants in the PV
coupling scheme shifts the position of a peak from the the very
forward angle for $g_{K^*N\Theta}=1.32$, $g_{K_1 N\Theta}=-0.1$ to
the very backward angle for $g_{K^*N\Theta}=-1.32$, $g_{K_1
N\Theta}=0.1$ . It is worth noting that the threshold behaviors of
the Born terms of these two processes given in Figs.
\ref{fig:fig04}, \ref{fig:fig05} and \ref{fig:fig06} show a close
similarity to those of $\gamma n\to \pi^- p$ and $\gamma p\to
\pi^0 p$ near threshold found in Refs. \cite{hans,fuch}, as
mentioned before.

\section{photoproduction for the negative parity $\Theta^+$}

We now turn to the case of $\Theta^+$ photoproduction when it has the negative
parity.
The electromagnetic coupling vertex of the negative  parity $\Theta^+$ baryon is given by,
\begin{eqnarray}\label{nega-em}
\mathcal{L}_{\gamma\Theta\Theta} = - \bar{\Theta}\gamma_5 \left[ Q_{\Theta}\gamma^\mu
- \frac{\kappa_\Theta^{}}{2M_\Theta} \sigma_{\mu\nu} \partial^\nu
  \right]\gamma_5 \Theta A^\mu.
\end{eqnarray}
The interaction Lagrangians of the negative parity $\Theta^+$ for
the PS and PV couplings are of the forms;
\begin{eqnarray}\label{nega-cpl-00}
&&\mathcal{L}^{PS}_{KN\Theta} = -i g_{KN\Theta}^{}\bar{\Theta}\, N\, K,\nonumber\\
&&\mathcal{L}^{PV}_{KN\Theta}
= - \frac{g_{KN\Theta}}{M-M_{\Theta}} \bar{\Theta}\gamma^{\mu}\, N\,
\partial_{\mu}K,
\end{eqnarray}
which are equivalent to each other for the free baryons. It must be noted, however,
that they are slightly different from each other when, reduced to
the non-relativistic spinor forms at threshold, i.e.,
\begin{eqnarray}\label{nega01}
&&\mathcal{L}^{PS}_{KN\Theta}=-ig_{KN\Theta}\,\chi^\dag_{\Theta}\,\chi_{N}\,K+\cdots~,\nonumber\\
&&\mathcal{L}^{PV}_{KN\Theta}
=-ig_{KN\Theta}\frac{m_{K}}{M-M_{\Theta}}\chi^\dag_{\Theta}\,\chi_{N}\,K+\cdots~.
\end{eqnarray}

The difference is by the factor $\frac{m_{K}}{M_{\Theta}-M}\simeq$
0.85, which makes the PV coupling version somewhat smaller than
the PS one by the factor of 0.85. The Born amplitude
for the negative parity $\Theta^+$ photoproduction can be derived by
using the Lagrangians in Eqs.(\ref{nega-em}) and
(\ref{nega-cpl-00}) for the coupling vertices relevant to the
interaction diagrams shown in Fig. \ref{fig:fig01}. This leads to
the replacement of the final state $\bar{u}_{\Theta}$ $\to$
$-\bar{u}_{\Theta}\gamma_5$ at every $KN\Theta$ vertex and
the u-channel  propagator of $\Theta^+$, $S_{F}(p^{\prime}-k)$ $\to$
$-\gamma_5 S_{F}(p^{\prime}-k)\gamma_5$ in the amplitude given by
Eq.(\ref{00}), i.e.,
\begin{eqnarray}\label{nega-pv-01}
{\cal M}_{Born}={\cal M}_{PV-pole}+{\cal M}_{KR}+{\cal M}_{c}\,,
\end{eqnarray}
where
\begin{eqnarray}\label{nega00}
&&{\cal M}_{PV-pole}\nonumber\\&=&\frac{eg_{KN\Theta}}{M-M_{\Theta}}\bar{u}_{\Theta}(p^{\prime})\biggl\{
-F_{1}(s)/\kern-5pt{q}\frac{(/\kern-6pt{p}+/\kern-5pt{k}+M)}{s-M^{2}}\,
[Q_{N}\,/\kern-5pt{\epsilon} +i\frac{\kappa_{N}}{2M}
\sigma^{\mu\nu}\epsilon_{\mu}k_{\nu}]
+[-Q_{\Theta}\,/\kern-5pt{\epsilon}
+i\frac{\kappa_{\Theta}}{2M_{\Theta}}\sigma^{\mu\nu}\epsilon_{\mu}k_{\nu}]\nonumber\\&&\times
 \frac{(/\kern-6pt{p^{\prime}}-/\kern-5pt{k}
+M_{\Theta})}{u-{{M_{\Theta}}^{2}}}/\kern-5pt{q}F_{2}(u)
-Q_K F_{3}(t)\frac{(M-M_{\Theta})(2q-k)\cdot\epsilon}{t-m_{K}^{2}}\,
\biggr\}\,u_{N}(p)\,,
\end{eqnarray}
\begin{eqnarray}\label{nega-kr-01}
{\cal M}_{KR}=\frac{eg_{KN\Theta}}{M-M_{\Theta}}\bar{u}_{\Theta}(p^{\prime})
\biggl\{\,F_{1}(s)\,Q_N- Q_{\Theta}\,F_{2}(u)
\biggr\}\,/\kern-5pt{\epsilon}\,u_{N}(p)\,,
\end{eqnarray}
\begin{eqnarray}\label{nega-oh-01}
{\cal M}_{c}&=&eg_{KN\Theta}\,\bar{u}_{\Theta}(p^{\prime})\biggl\{
(F_{1}(s)-\widehat{F})\,Q_{N}\frac{(2p+k)\cdot \epsilon}{s-M^{2}}\,
+Q_{\Theta}\,(F_{2}(u)-\widehat{F})\frac{(2p^{\prime}-k)\cdot
\epsilon}{u-{{M_{\Theta}}^{2}}}\nonumber\\&&
+Q_K(F_{3}(t)-\widehat{F})\frac{(2q-k)\cdot \epsilon}{t-m_{K}^{2}}
\biggr\} u_N(p)\,.
\end{eqnarray}
By the similar procedure,  the PS coupling Born amplitude 
is given by,
\begin{eqnarray}\label{nega-ps-00}
{\cal M}_{Born}={\cal M}_{PS-pole}+{\cal M}_c\,,
\end{eqnarray}
where the contact interaction term ${\cal M}_c$ is given by
the same equation, Eq.(\ref{nega-oh-01}), and
\begin{eqnarray}
&&{\cal M}_{PS-pole}\nonumber\\&=&eg_{KN\Theta}\,\bar{u}_{\Theta}(p^{\prime})
\biggl\{
-F_{1}(s)\,\frac{(/\kern-6pt{p}+/\kern-5pt{k}+M)}{s-M^{2}}\,
[\,Q_{N}\,/\kern-5pt{\epsilon} +i\frac{\kappa_{N}}{2M}
\sigma^{\mu\nu}\epsilon_{\mu}k_{\nu}\,]
+[\,-Q_{\Theta}\,/\kern-5pt{\epsilon}
+i\frac{\kappa_{\Theta}}{2M_{\Theta}}\sigma^{\mu\nu}\epsilon_{\mu}k_{\nu}\,]\nonumber\\&&\times
 \frac{(/\kern-6pt{p^{\prime}}-/\kern-5pt{k}
+M_{\Theta})}{u-{{M_{\Theta}}^{2}}}\,F_{2}(u)
-Q_K F_{3}(t)\frac{(2q-k)\cdot\epsilon}{t-m_{K}^{2}}\,
\biggr\}\,u_{N}(p).
\end{eqnarray}
For an application of the $K^*$ and $K_1$ exchanges in the
t-channel, we use the transition amplitudes of the positive parity $\Theta^+$ cases,
i.e. Eqs.(\ref{posi-vec}) and (\ref{posi-ax}), replacing $\bar{u}_{\Theta}$
by $-\bar{u}_{\Theta}\gamma_5$ in the $K^*N\Theta$ and
$K_1 N\Theta$ vertices.

\begin{figure}[tr]
\centering
\epsfig{file=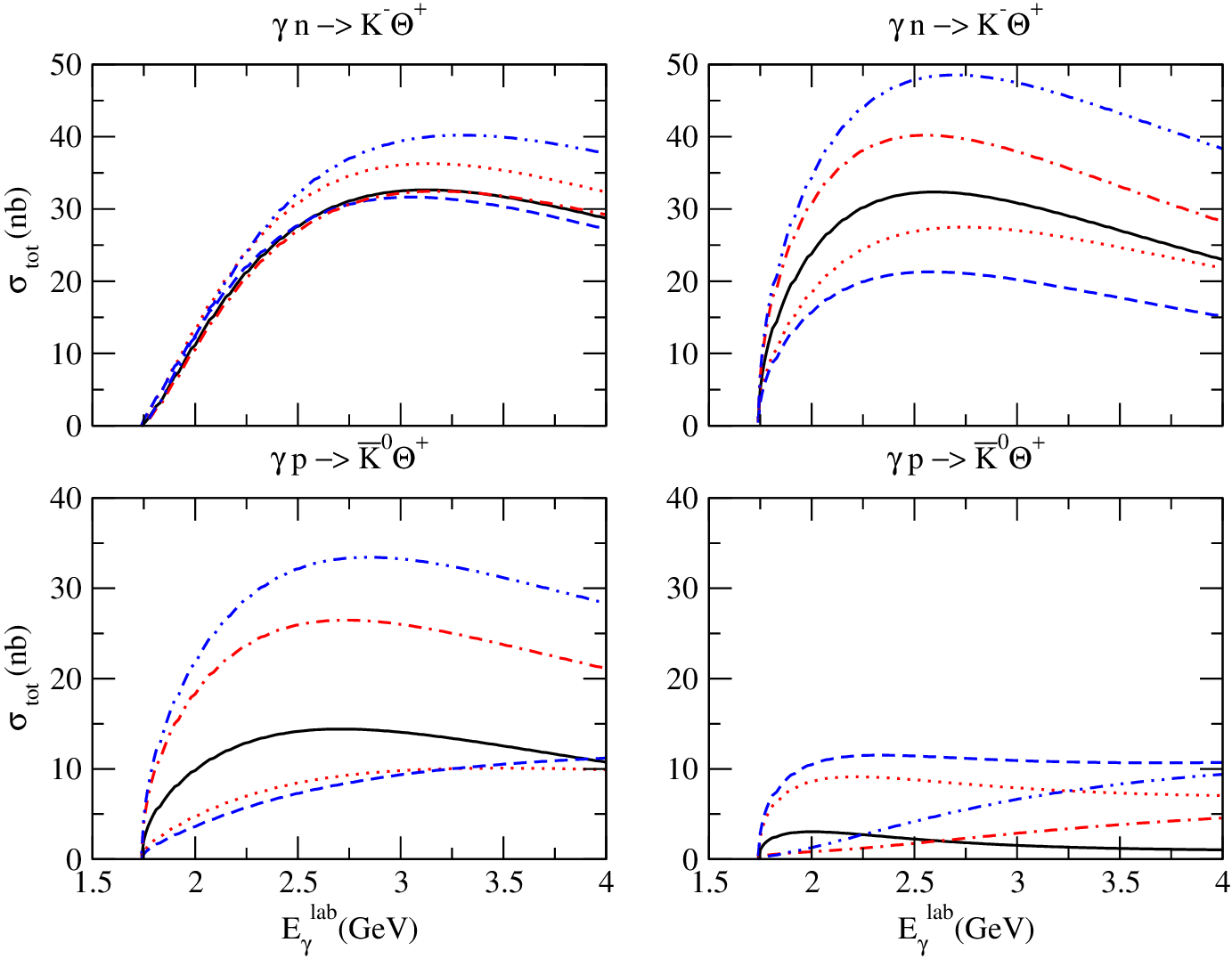, width=0.7\hsize}
\caption{Cross
sections for $\gamma n \to K^- \Theta^+$ of PV(upper left) and
PS scheme(upper right). Cross sections for $\gamma p \to \bar{K}^0
\Theta^+$ of PV(lower left) and PS scheme(lower right) when the
$\Theta^+$ has negative parity.
$\kappa_\Theta=0$ in any cases for all panels.
The solid lines are the contribution of the Born amplitude with
$g_{KN\Theta}=0.3$, $g_{K^*N\Theta}=0$. The dotted lines are the sum
of the Born amplitude and $K^*$ with $g_{K^*N\Theta}=0.18$. The
dot-dashed lines the sum of the Born amplitude and $K^*$ with
$g_{K^*N\Theta} = -0.18$. The dashed lines are the sum in total of
the Born terms, $K^*$ and $K_1$ with $g_{K^*N\Theta}= 0.18$,
$g_{K_1 N\Theta}= -0.1$ for $\gamma n\to K^{-}\Theta^+$ and
$g_{K_1 N\Theta}= -0.16$ for $\gamma p\to \bar{K}^{0}\Theta^+$
respectively. The dot-dot-dashed lines are the sum in total of the
Born, $K^*$ and $K_1$ with $g_{K^*N\Theta}= -0.18$, $g_{K_1
N\Theta}= +0.1$ for $\gamma n\to K^-\Theta^+$ and $g_{K_1
N\Theta}= +0.16$ for $\gamma p\to \bar{K}^{0}\Theta^+$
respectively. }
\label{fig:fig071}
\end{figure}

The process for the negative parity $\Theta^+$ production was
considered in Refs. \cite{nam,cmko02,ysoh,zhao} and found to have
smaller cross section than the positive-parity $\Theta^+$
production. In Fig. \ref{fig:fig071}, the total cross sections
are shown for both processes with $g_{KN\Theta}=0.3$ taken from
the decay width $\Gamma_{\Theta} \simeq 5$ MeV. For the $K^*$ and
$K_1$ coupling constants, we use $g_{K^*N\Theta}=0.18$, keeping
the ratio $g_{K^*N\Theta}/g_{KN\Theta}=0.6$.
The coupling constant $g_{K_1 N\Theta}$ is determined from the
assumption that the ratio $g_{K^*K\gamma}g_{K^*
p\Lambda(1405)}/g_{K_1 K\gamma}g_{K_1 p\Lambda(1405)}= -0.7$
extracted from Ref. \cite{crji} holds for the present coupling
ratio $g_{K^*K\gamma}g_{K^* N\Theta}/g_{K_1 K\gamma}g_{K_1
N\Theta}$ as well. In the figures, the role of $K_1$ is
appreciable in the negative parity $\Theta^+$ and the cross
sections are sensitive to the sign of $K^*$ and $K_1$ coupling
constants. This is analogous to the $K^*$ dominance in the
positive parity production, since the parities of $K^*$ and $K_1$
are opposite to each other. Depending on the signs of the $K^*$
and $K_1$ coupling constants, the cross sections for $\gamma n\to
K^-\Theta^+$ are around 30 nb for PV, and 20 $\sim$ 50 nb for PS
schemes at $E_{\gamma}=2.5$ GeV respectively. While for $\gamma
p\to\bar{K}^0\Theta^+$, the cross sections are about 7 $\sim$ 33
nb  in the PV, and 2 $\sim$ 12 nb in the PS schemes. From these
figures we find that the reaction $\gamma n \to K^- \Theta^+$ is
still dominant over the reaction $\gamma p \to \bar{K}^0 \Theta^+$
in the case of the negative parity $\Theta^+$ as well. It should
be noted that the inclusion of magnetic moment $\kappa_\Theta$
could give an additional contribution to the cross sections.

In comparison with the cross sections of the positive parity
$\Theta^+$ production in Fig. \ref{fig:fig04}, the cross sections
in the case of negative parity are suppressed roughly by an order
of magnitude.
This is consistent with the previous calculations presented in
Refs. \cite{cmko01,ysoh}. The reason for the suppression is mainly
because the adopted coupling constant $g_{KN\Theta}=0.3$ taken
from $\Gamma_{\Theta}\simeq 5$ MeV is smaller by a factor of
$\frac{1}{7}$ than that of positive parity. This reduction is of
course reflected in the suppression of cross sections roughly by
an order of magnitude smaller than the existing SAPHIR
experimental cross section. Thus, if we trust the existing
SAPHIR data, then we may well doubt the
possibility of negative parity state of $\Theta^+$.  However, there
exists a rather large uncertainty in the present measurement of
the decay width of the $\Theta^+$. Also, as we demonstrated in
the positive parity case, the cross sections calculated in the
framework of hadron models are largely dependent on the gauge
prescription as well as the cutoff $\Lambda$.
Moreover, let us consider the decay width of the transition,
$\Theta^+(\frac{1}{2}^{\pm}) \to K N(\frac{1}{2}^{+})$ with both
parities retained; i.e.,
\begin{eqnarray}\label{decay}
\Gamma_{\Theta(\frac{1}{2}^{\pm})}
=\frac{g^2_{KN\Theta}}{2\pi}\frac{|\bm{q}|}{M_{\Theta}}(\sqrt{M^2+|\bm{q}|^2}\mp
M\,),
\end{eqnarray}
and suppose that the coupling constant $g_{KN\Theta}$ is $a\;
priori$ given and the kaon momentum $|\bm{q}|$ in the $\Theta^+$
rest frame is small. Then, the Eq.(\ref{decay}) implies that the
width $\Gamma_\Theta$ near threshold would be small for the
positive parity $\Theta^+$ by the subtraction of nucleon mass from
its energy, and vice versa for the negative parity. This means
that the decay of the positive parity is kinematically forbidden
for small momentum and initially a negative parity state of
$\Theta^+$ is more likely to decay to the final nucleon and kaon.
%
In this respect, analyzing only the
total cross sections does not seem to provide a decisive
conclusion on the parity of the $\Theta^+$. We thus analyze
further these processes by presenting the angular distributions.
\begin{figure}[tr]
\centering \epsfig{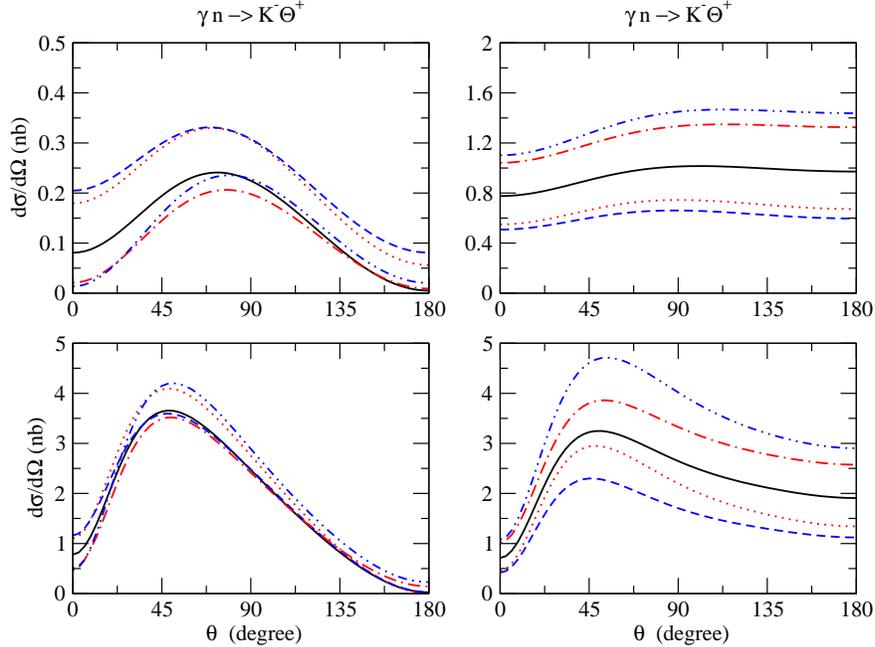}
\caption{Angular distributions for $\gamma n \to K^- \Theta^+$ at
$E_\gamma=1.8$ GeV(upper left), $E_\gamma=2.5$ GeV(lower left) of
PV and $E_\gamma=1.8$ GeV(upper right), $E_\gamma=2.5$ GeV(lower
right) of PS scheme with the negative parity $\Theta^+$. The
notations are the same as in Fig. \ref{fig:fig071}}
\label{fig:fig09}
\end{figure}
\begin{figure}[tr]
\centering \epsfig{file=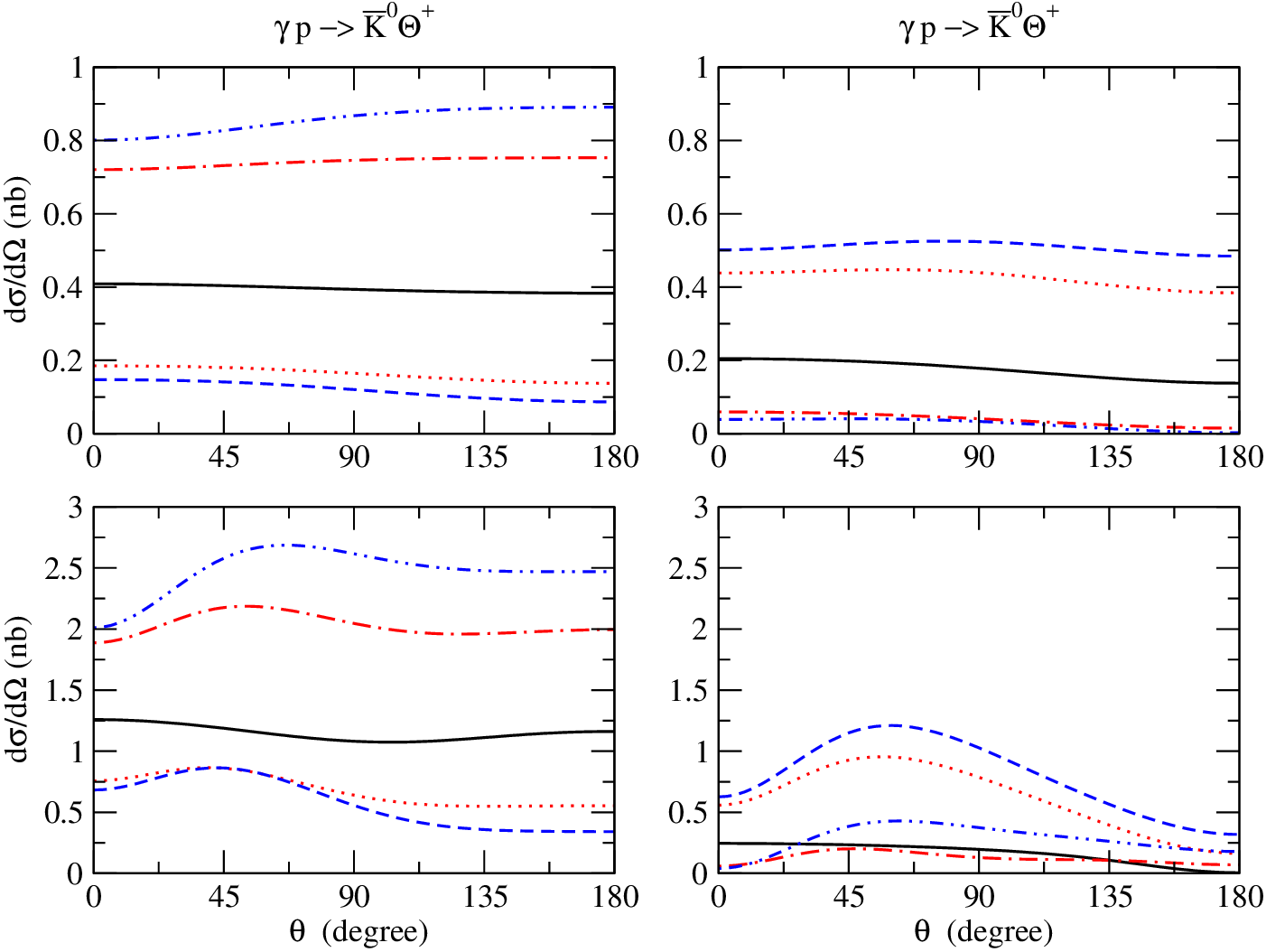, width=0.7\hsize}
\caption{Angular distributions for $\gamma p \to \bar{K}^0
\Theta^+$ at $E_\gamma=1.8$ GeV(upper left), $E_\gamma=2.5$
GeV(lower left) of PV and $E_\gamma=1.8$ GeV(upper right),
$E_\gamma=2.5$ GeV(lower right) of PS scheme with the negative
parity $\Theta^+$. The notations are the same as in Fig.
\ref{fig:fig071}}
\label{fig:fig10}
\end{figure}

The angular distributions for $\gamma n\to K^-\Theta^+$ and
$\gamma p\to \bar{K}^0\Theta^+$ are displayed in Figs.
\ref{fig:fig09} and \ref{fig:fig10}, respectively. By the parity
and angular momentum conservation, the produced kaon is
anticipated to be in the p-wave state near threshold when the
produced $\Theta^+$ has negative parity. Such a feature is well
reproduced in $\gamma n\to K^-\Theta^+$, whereas it is less clear
for $\gamma p\to \bar{K}^0\Theta^+$, regardless of the coupling
schemes of the $KN\Theta^+$ interaction. Note that the scales of
the cross sections of $E_\gamma$=1.8 GeV in the two schemes are
different in Fig. \ref{fig:fig09}. In particular, at the photon
energy $E_\gamma=2.5$ GeV we observe a forward peak due to a
coherent interference of the Born terms with $K^*$ and $K_1$ right
around 45$^\circ$ both in the two schemes in Fig. \ref{fig:fig09}.
The coherent peak of the Born terms around 45$^\circ$ is
understood by the t-channel kaon pole dominance. In the case of
$\gamma p\to \bar{K}^0\Theta^+$ presented in Fig. \ref{fig:fig10},
the apparent flat curves of the Born contribution may be due to
the small u-channel Born contribution weakened by the small
coupling constant $g_{KN\Theta}$. Therefore, the development of
the angular distribution of the cross section in this process
comes from the t-channel $K^*$ and $K_1$ contributions as the
photon energy increases.

Before closing this section, it should be remarked that the
angular distributions of $\gamma n\to K^-\Theta^+$ and $\gamma
p\to \bar{K}^0\Theta^+$ near threshold in particular show a clear
distinction between two opposite parities of the $\Theta^+$ baryon
and they are given in a rather model-independent way. As we have
demonstrated up to this point, near threshold where the orbital
excitations of kaon other than $L=0$ or $1$ are suppressed, the
conservation of parity and angular momentum imposes a specified
form on the shape of angular distribution of $\gamma N \to K
\Theta^+$, depending on what parity state of the $\Theta^+$ is.
Therefore, the observation of the reaction near threshold can
provide an unambiguous way to clarify the parity of the
$\Theta^+$. The importance of using this sort of conservation laws
near threshold was also emphasized in Ref. \cite{thom}, but for
the different reaction $pp\to \Sigma^+\Theta^+$. The reaction they
suggested instead of the $\Theta^+$ photoproduction takes the
advantage of giving more tight condition on the parity and angular
momentum at the initial $pp$ state. However, the photoproduction
of $\Theta^+$ has already been observed and seems to be more
available for the present experiment than the reaction suggested
in Ref. \cite{thom}.

\section{Summary and Discussion}

We investigated the possibility of using photon induced $\Theta^+$
production, $\gamma n\to K^-\Theta^+$ and $\gamma p\to
\bar{K}^0\Theta^+$, to discriminate the parity of the $\Theta^+$
baryon. The processes are calculated for two possible parity
states of the $\Theta^+$ baryon using the hadron model where the
interaction of the $KN\Theta$ vertex is considered both in the PV
and PS coupling schemes. We employ the broader basis of
prescription for the gauge invariance based on the Ohta and
Haberzettl methods, as discussed in the study of $K^+\Lambda$ and
$K^+\Sigma$ photoproductions \cite{ohta,habe,habe01}. The results
for the total and differential cross sections are to a large
extent different from those of previous calculations
\cite{nam,cmko01,ysoh,zhao}, indicating that there is a wide
theoretical uncertainty associated with the different assumptions
on the gauge prescription. We may summarize the differences as
follows.
(i) With the decay width 5 MeV, the cross sections of the positive
parity $\Theta^+$ from the neutron and proton target are
comparable to the present SAPHIR data, whereas the cross sections
of the negative parity are found to be only tens of nb.
Nevertheless, the uncertainty in the experimental data are now
under discussion and if the cross sections are indeed an order of
magnitude smaller, as presented in Refs. \cite{clos,hosa2}, than
the published SAPHIR data \cite{saph}, then the results of the
present work are likely to support the photoproduction by the
negative parity $\Theta^+$ even with the coupling constant
$g_{KN\Theta}=0.3$.
Our results also show that the cross section of the $\Theta^+$
production from the neutron is on the whole larger than that of
$\Theta^+$ production from the proton.
(ii) Using the empirical ratio of
$\frac{g_{K^*K\gamma}g_{K^*N\Lambda}}{g_{K_1K\gamma}g_{K_1
N\Lambda}}$ extracted from $K^+ \Lambda$ electromagnetic
production for the determination of the coupling constants,
$g_{K^*N\Theta}$ and $g_{K_1 N\Theta}$, we obtain the $K^*$ and
$K_1$ contributions and find that the $K^*$ contribution is in
general important and $K_1$ contribution to the negative
$\Theta^+$ parity is not negligible either. However, these
contributions are not so much dominant over the Born contribution
as claimed in Ref. \cite{ysoh}. This point is supported by the
generally known fact that the contributions of vector meson
$\rho(\omega)$ are about 10 $\%$ of the Born contributions at best
to the threshold amplitudes  of  the reactions $\gamma n\to \pi^-
p$ and $\gamma p\to \pi^0 p$ \cite{tiat}.
(iii) Finally, we find that the angular distributions of the
production processes for the two opposite parity states are less
dependent on the model parameters and distinct from each other.
Therefore, we suggest that the observation of angular distribution
in the photoproduction process can serve as a more useful tool to
distinguish the parity of the $\Theta^+$ baryon as compared to the
measurement of total cross sections only.

\acknowledgments

This work was supported in part by 2003 Hankuk Aviation university
Faculty Research Grant and in part by a grant from the U.S.
Department of Energy (DE-FG02-96ER 40947).


\begin{thebibliography}{10}

\bibitem{leps}
\mbox{LEPS Collaboration,} T.~Nakano {\em et~al.\/},
  Phys. Rev. Lett. {\bf 91}, 012002 (2003).

\bibitem{saph}
\mbox{SAPHIR Collaboration,} J.~Barth {\em et~al.\/},
  Phys. Lett. B {\bf 572}, 127 (2003);~hep-ex/0307083.

\bibitem{clasi}
\mbox{CLAS Collaboration I,} S.~Stepanyan {\em et~al.\/},
  Phys. Rev. Lett. {\bf 91}, 252001 (2003).

\bibitem{clasii}
\mbox{CLAS Collaboration II,} V.~Kubarovsky {\em et~al.\/},
  Phys. Rev. Lett. {\bf 92}, 032001 (2004).

\bibitem{herm}
\mbox{HERMES Collaboration,} A.~Airapetian {\em et~al.\/},
  Phys. Lett. {\bf B 585}, 213 (2004).

\bibitem{dian}
V.~V.~Barmin {\em et~al.\/} (\mbox{DIANA Collaboration}),
  Phys. Atom. Nucl. {\bf 66}, 1715 (2003).

\bibitem{asra}
A.~E. Asratyan, A.~G. Dolgolenko, and M.~A. Kubantsev,
  hep-ex/0309042.


\bibitem{svd}
\mbox{SVD Collaboration,} A.~Aleev {\em et~al.\/},
  hep-ex/0401024.

\bibitem{diak}
D.~Diakonov, V.~Petrov, and M.~Polyakov,
  Z. Phys. A {\bf 359}, 305 (1997).


\bibitem{weig}
H.~Weigel,
  Eur. Phys. J. A {\bf 2}, 391 (1998).

\bibitem{chem}
M.~Chemtob,
  Nucl. Phys. {\bf B256}, 600 (1985).


\bibitem{pras}
M.~Prasza{\l}owicz,
  Phys. Lett. B {\bf 575}, 234 (2003).

\bibitem{jaff}
R.~L. Jaffe and F.~Wilczek,
   Phys. Rev. Lett. {\bf 91}, 232003 (2003).

\bibitem{jaff01}
R.~L. Jaffe,
  Phys. Rev. D {\bf 15}, 267 (1977);~
  Phys. Rev. D {\bf 15}, 281 (1977).

\bibitem{cheu}
K.~Cheung
  hep-ph/0308176.

\bibitem{stan}
Fl.~Stancu and D.~O. Riska,
  Phys. Lett. B {\bf 575}, 242 (2003).

\bibitem{carl}
C.~E. Carlson, C.~D. Carone, H.~J. Kwee, and V.~Nazaryan,
  Phys. Lett. B {\bf 573}, 101 (2003);~
  Phys. Lett. B {\bf 579}, 52 (2004).

\bibitem{golz}
L.~Ya. Golzman,
  Phys. Lett. B {\bf 575}, 18 (2003); ~hep-ph/0309092.

\bibitem{will}
R.~A. Williams and P.~Gu{\`{e}}ye,
  nucl-th/0308058.


\bibitem{karl}
M.~karliner and H.~J. Lipkin,
   Phys. Lett. B {\bf 575}, 249 (2003).


\bibitem{shur}
E.~Shuryak and I.~Zahed,
  hep-ph/0310270.

\bibitem{wall}
H.~Walliser and V.~B. Kopeliovich,
  J. Exp. Theor. Phys. {\bf 97}, 433 (2003).

\bibitem{jenn}
B.~K. Jennings and K.~Maltman,
  hep-ph/0308286.

\bibitem{bori}
D.~Borisyuk, M.~Faber, and A.~Kobushkin,
  hep-ph/0307370.


\bibitem{itzh}
N.~Itzhaki, I.~R. Klebanov, P.~Ouyang, and L.~Rastelli,
  hep-ph/0309305.

\bibitem{wu}
B.~Wu and B.-Q. Ma,
  hep-ph/0311331.

\bibitem{zhu}
S.-L. Zhu,
  Phys. Rev. Lett. {\bf 91}, 232002 (2003).

\bibitem{math}
R.~D. Matheus, F.~S. Navarra, M.~Nielsen,
R.~da Silva, and S.~H. Lee,
  Phys. Lett. B {\bf 578}, 323 (2004).

\bibitem{sugi}
J.~Sugiyama, T.~Doi, and M.~Oka,
  Phys. Lett. B {\bf 581}, 167 (2004).

\bibitem{hosa}
A.~Hosaka,
  Phys. Lett. B {\bf 571}, 55 (2003).


\bibitem{cohe}
T.~D. Cohen,
  Phys. Lett. B {\bf 581}, 175 (2004)
  ; T.~D. Cohen and R.~F. Lebed,
  Phys. Lett. B {\bf 578}, 150 (2004).


\bibitem{csik}
F.~Csikor, Z.~Fodor, S.~D. Katz, and T.~G. Kov{\'{a}}cs,
  hep-lat/0309090.

\bibitem{sasa}
S.~Sasaki,
  hep-lat/0310014.


\bibitem{wybo}
B.~G. Wybourne,
  hep-ph/0307170.

\bibitem{bijk}
R.~Bijker, M.~M. Giannini, and E.~Santopinto
  hep-ph/0310281.



\bibitem{rand}
J.~Randrup,
  Phys. Rev. C {\bf 68}, 031903 (2003).

\bibitem{chen}
L.~W. Chen, V.~Greco, C.~M. Ko, S.~H. Lee, and W.~Liu
  nucl-th/0308006.


\bibitem{nam}
S.-I. Nam, A.~Hosaka, and H.-C. Kim,
  Phys. Lett. B {\bf 579}, 43 (2004).


\bibitem{cmko01}
W.~Liu and C.~M. Ko,
  Phys. Rev C {\bf 68}, 045203 (2003);~
  nucl-th/0309023.

\bibitem{cmko02}
W.~Liu and C.~M. Ko,
  Phys. Rev. C {\bf 69}, 025202 (2004).

\bibitem{ysoh}
Y.~S. Oh, H.~C. Kim, and S.-H. Lee,
  Phys. Rev. D {\bf 69}, 014009 (2004).

\bibitem{zhao}
Q.~Zhao and J.~S. Al-Khalili,
  hep-ph/0310350.

\bibitem{naka}
K.~Nakayama and K.~Tsushima,
  Phys. Lett. B {\bf 583}, 269 (2004).

\bibitem{jans01}
S.~Janssen, J.~Ryckebusch, D.~Debruyne, and T.~Van Cauteren,
  Phys. Rev. C {\bf 66}, 035202 (2002);~ Phys. Rev. C {\bf 65}, 015201 (2001).

\bibitem{jans02}
S.~Janssen, J.~Ryckebusch, W.~Van Nespen, D.~Debruyne, and T.~Van Cauteren,
    Eur. Phys. J. A {\bf 9}, 115 (2000).

\bibitem{benn}
C.~Bennhold, T.~Mart, A.~Waluyo, H.~Haberzettl, G.~Penner,
T.~Feuster, and U.~Mosel,
  nucl-th/9901066.


\bibitem{mart}
T.~Mart, S.~Sumowidagdo, C.~Bennhold, and H.~Haberzettl,
  nucl-th/0002036.


\bibitem{ohta}
K.~Ohta,
  Phys. Rev. C {\bf 40}, 1335 (1989);~
  Phys. Rev. C {\bf 46}, 2519 (1992).

\bibitem{habe}
H.~Haberzettl,
  Phys. Rev. C {\bf 56}, 2041 (1997).
H.~Haberzettl {\em et~al.\/},
  nucl-th/9804051.
H.~Haberzettl,
  Phys. Rev. C {\bf 62}, 034605 (2000).

\bibitem{habe01}
H.~Haberzettl, C.~Bennhold, T.~Mart, and T.~Feuster,
  Phys. Rev. C {\bf 58}, R40 (1998).

\bibitem{davi}
R.~M. Davidson and R.~Workman,
  Phys. Rev. C {\bf 63}, 025210 (2001);~
  Phys. Rev. C {\bf 63}, 058201 (2001).

\bibitem{nuss}
S.~Nussinov,
  hep-ph/0307357.

\bibitem{arnd}
R.~A. Arndt, I.~I. Strakovsky, and R.~L. Workman,
  nucl-th/0311030.

\bibitem{clos}
F.~E. Close and Q.~Zhao,
  hep-ph/0403159.

\bibitem{ysoh01}
Y.~S. Oh, H.~C. Kim, and S.~H. Lee,
  hep-ph/0310117.



\bibitem{diak01}
D.~Diakonov and V.~Petrov,
  hep-ph/0310212.

\bibitem{tiat}
D.~Drechsel and L.~Tiator,
  J. Phys. G: Nucl. Part. Phys. {\bf 18}, 449 (1992).



\bibitem{hans}
O.~Hanstein, D.~Drechsel, and L.~Tiator,
  nucl-th/9709067;
  Nucl. Phys. A {\bf 632}, 561 (1998).

\bibitem{fuch}
M.~Fuchs {\em et~al.\/}, Phys. Lett. B {\bf 368}, 20 (1996).


\bibitem{poly03}
M.~V. Polyakov and A.~Rathke,
  hep-ph/0303138;
  Eur. phys. J. A {\bf 18}, 691 (2003).

\bibitem{pdg}
http://pdg.lbl.gov/pdg.html

\bibitem{hagl}
K.~Haglin,
  Phys. Rev. C {\bf 50}, 1688 (1994).


\bibitem{crji}
R.~A. Williams, C.-R. Ji, and S.~R. Cotanch,
  Phys. Rev. C {\bf 46}, 1617 (1992).


\bibitem{thom}
A.~W. Thomas, K.~Hicks, and A.~Hosaka,
  hep-ph/0312083;
  Prog. Theor. Phys. {\bf 111}, 291 (2004).

\bibitem{hosa2}
S.-I. Nam, A.~Hosaka, and H.-C. Kim,
  hep-ph/0403009;


\end{thebibliography}
\end{document}